# Price risk aversion vs payoff risk aversion: a gender comparison through a laboratory experiment

## Ali Zeytoon-Nejad
*School of Business, Wake Forest University, Winston-Salem, North Carolina, USA*

## Abstract

**Purpose** – This paper explores gender differences in two distinct forms of risk aversion—Payoff Risk Aversion (PaRA) and Price Risk Aversion (PrRA)—in order to provide a more nuanced understanding of how men and women respond to different types of economic uncertainty.

**Design/methodology/approach** – The study employs a laboratory experiment using Multiple-Choice-List (MCL) risk-elicitation tasks based on both Direct Utility Function (DUF) and Indirect Utility Function (IUF) frameworks. These tasks present stochastic payoffs and stochastic prices, respectively. The analysis uses statistical hypothesis testing to compare gender-specific responses across three experimental designs.

**Findings** – The key results of the study indicate that women typically exhibit higher degrees of PaRA than men, which is a consistent finding with the mainstream literature. However, remarkably, the results from all the three indirect MCL designs show that women typically exhibit lower degrees of PrRA than men, and this result is robust across different MCL designs. The paper also introduces an "irrationality gap" as the difference between PaRA and PrRA and explores the size of the irrationality gap within either gender group, finding it larger and statistically significant for men, while smaller and statistically insignificant for women.

**Originality/value** – This study is the first to distinguish between PaRA and PrRA in a gender comparison, using experimentally validated methods. It provides new behavioral insights into the nature of gender-specific risk preferences and introduces the irrationality gap as a novel concept with implications for understanding financial decision-making and the design of gender-sensitive economic policies.

**Paper type** Research article

## 1. Introduction

Scholars have identified gender differences in a number of various domains, such as investment decisions, insurance preferences, retirement plans, and choices and assignments within the labor market. There is considerable empirical evidence suggesting that women are generally more (payoff) risk-averse than men; therefore, they are more likely to make lower-risk/lower-return choices than men. The main purpose of the present paper is to delve into gender differences in risk aversion, examining not only "Payoff" Risk Aversion (PaRA) but also "Price" Risk Aversion (PrRA) [1], as introduced by Zeytoon Nejad Moosavian *et al.* (2020). Accordingly, the present study uses the experimental data collected by Zeytoon Nejad Moosavian *et al.* (2020), and builds upon that work, which used various Multiple-Choice-List (MCL) designs (aka Multiple-Price-List (MPL) designs).

MCL designs have emerged as popular elicitation methodologies in experimental economics for assessing risk attitudes and quantifying the extent of risk aversion within the controlled environment of experimental laboratories employing non-interactive setups. In Zeytoon Nejad Moosavian *et al.* (2020), risk elicitation designs were deliberately calibrated in a way that, given Expected Utility Theory (EUT) and Duality Theory (DT), each design must

*Funding statement:* This study has been funded by the Wake Forest University School of Business.

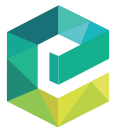

elicit the same degree of risk aversion exhibited by a given rational economic agent, although the designs differ in form, i.e. in terms of their approaches (i.e. DUF vs. IUF) and their elicitation designs (i.e. three MCL designs). They showed that the sense of PrRA is stronger than the sense of PaRA in humans in general, meaning that PaRA and PrRA are empirically distinct preference parameters. The present study attempts to break down the results into different genders in the hope that a more holistic understanding of gender differences with respect to different types of risk aversion can be developed.

Accordingly, the primary research questions of this paper are as follows: Is there any significant difference in the degrees of risk aversion with respect to stochastic prices and stochastic payoffs across different genders? What type of risk attitudes do men and women exhibit under uncertainty about the prices of goods they are to buy? If risk aversion is the dominant risk attitude, to what extent are men and women risk averse with respect to stochastic prices and stochastic payoffs? How much are men's and women's risk premiums over the lotteries defined? (In other words, how large of a premium are men and women willing to pay to set prices fixed ex-ante?) Finally, what are some psychological and behavioral explanations for the differences observed in the degrees of risk aversion elicited through the two distinct approaches described above across gender? This study contributes to a growing literature on gender differences in the behavior of economic agents under risk.

A series of studies in the respective literature align in the assertion that women tend to be more risk-averse than men. For example, Dohmen *et al.* (2011) and Sutter *et al.* (2013) find women to be less willing to take risks than men using large representative surveys and experiments. Eckel and Grossman (2008a, b) also support this view through various experimental methodologies. In contrast, Sarin and Wieland (2016), Harrison *et al.* (2007), and Hillesland (2019) argue for no statistically significant gender differences in risk aversion. Croson and Gneezy (2009) propose that while women are generally more risk-averse, exceptions exist, especially among professionals. Schubert *et al.* (1999) and Holt and Laury (2002) illustrate context-dependent gender differences, where risk aversion levels may vary based on decision framing and context. While prior studies have primarily focused on aggregate gender differences in risk attitudes, this study offers a novel dimension by examining how framing and context (payoffs vs. prices) interact with gender.

According to Zeytoon Nejad Moosavian *et al.* (2020), the data collection procedure of this study adopted the frameworks of three popular MCL designs, including Holt and Laury (2002), Binswanger (1980), and Certainty-vs.-Uncertainty design (henceforth, the H&L, Bins., and CvU designs, respectively), and used all the three MCL designs (i.e. three contexts) in two versions (i.e. two approaches) – a version with a DUF approach and another with an IUF approach. Accordingly, the experimental design had a $3 \times 2$ design. For each of the six treatments, four independent sessions were conducted. The experimental subjects were 88 students studying at North Carolina State University from a wide range of disciplines who participated in the experiment using the experimental-economics software zTree on computers in the experimental economics laboratory of the Department of Economics at North Carolina State University. For the purpose of statistical hypothesis-testing, a wide range of relevant statistical tests, including Wilcoxon-Mann-Whitney test, Kolmogorov-Smirnov equality-of-distributions test, and Two-Sample *T* Test for paired data are employed.

The general pattern of results in the present paper indicates that women tend to display higher levels of Payoff Risk Aversion (PaRA) than men (which is consistent with a body of prior literature such as Eckel and Grossman (2008a, b), Dohmen *et al.* (2011), and Sutter *et al.* (2013)), while, conversely, they tend to exhibit lower levels of Price Risk Aversion (PrRA) compared to men. Overall, the findings of the study reflect the importance of the influence of decision framing and contextual differences (which is consistent with another body of prior literature such as Schubert *et al.* (1999), Holt and Laury (2002), and Croson and Gneezy (2009)), meaning that women are more risk-averse when it comes to stochastic payoffs while men are more risk-averse when it comes to stochastic prices.

In addition to documenting novel patterns of gender-based differences in risk preferences across payoff and price contexts, this paper also contributes to the theoretical literature by articulating a set of explanatory hypotheses grounded in established behavioral and psychological frameworks. Drawing from theories such as prospect theory, ambiguity aversion, the endowment effect, and gender-based socialization, the observed asymmetries between PaRA and PrRA are interpreted through the lens of differential cognitive framing, loss sensitivity, and decision consistency across genders. While the current study does not empirically test these mechanisms directly, it establishes a coherent theoretical foundation for future experimental inquiry and hypothesis-driven research in behavioral economics and gender studies. The findings of this study present numerous significant implications as well. For example, financial institutions and insurance companies can tailor their offerings to better align with gender-specific risk preferences, potentially promoting preference-matching efficiency and fostering greater inclusivity and customer satisfaction.

The rest of this paper is structured as follows: Section two delves into the experimental design and tasks, providing a brief overview of the data, variables, and the estimation strategy and procedures employed in this study. Section three reports and discusses the results of the study and estimations. In Section four, potential psychological and behavioral explanations for the observed disparities are reviewed and a coherent theoretical foundation for future experimental inquiry and hypothesis-driven research in behavioral economics is established. In Section five, the economic and business implications of the findings are introduced and discussed. In Section six, a conclusion will follow bringing the major findings together and proposing plans for future research.

## 2. Experimental design and procedures

To collect the dataset used in this study, Zeytoon Nejad Moosavian *et al.* (2020) employed a methodology akin to Holt and Laury (2002), categorizing risk attitudes into ten classifications. They designed lottery choice menus for three distinct Multiple-Choice-List (MCL) designs [2], including Holt and Laury (2002), Binswanger (1980), and Certainty-vs.-Uncertainty design (thereafter, the H&L, Bins., and CvU designs, respectively), deliberately calibrating them to ensure equivalence under the Expected Utility Theory (EUT). Constant Relative Risk Aversion (CRRA) was assumed for utility functional form due to its computational ease, theoretical support, robust predictions, and mathematical tractability [3].

To compare Price Risk Aversion (PrRA) with Payoff Risk Aversion (PaRA), two sets of theoretically equivalent menus were created for each MCL design, ensuring equivalence under Duality Theory (DT), introducing uncertainty about "prices" and "payoffs" [4]. This allowed the isolation of responses to uncertainties regarding "prices" and "payoffs" [5]. The experiments presented subjects with lotteries involving different odds, categorized into tasks with "payoff odds" and "price odds." In fact, there are two types of tasks: Tasks with "payoff odds" (in which payoff odds determine the payoff that the subjects will receive depending on their choices) and tasks with "price odds" (in which price odds determine the price that the subjects will buy widgets at). In the former type, the game of chance involves differing odds regarding whether the subjects receive a higher or lower "payoff." In the latter type, the game of chance involves differing odds regarding whether the subjects will buy at a higher or lower "price." In these tasks, the subjects buy and sell widgets. That is, they will buy widgets at some "uncertain buying prices," and will sell them to the experimenter at a "certain selling price." To determine their final payoffs, a six-sided die will eventually be rolled to select one of the six tasks for their payment at random [6].

The experimental design employed in this study builds on and extends the framework introduced in Zeytoon Nejad Moosavian *et al.* (2020), using three widely established MCL formats: (1) the Holt and Laury (2002) design with varying probabilities, (2) the Binswanger (1980) design with varying payoff magnitudes, and (3) a Certainty-versus-Uncertainty (CvU) design adapted from Balsa *et al.* (2015). Each of these was implemented in two theoretically equivalent versions: one based on stochastic payoffs (to elicit PaRA) using the Direct Utility Function (DUF), and the other based on stochastic prices (to elicit PrRA) using the Indirect Utility Function (IUF). Together, this yielded six total tasks per subject. Each MCL consisted of 10 binary choice rows—e.g. a safe vs. a risky lottery—where the probabilities or payoffs changed systematically across the list. The equivalence of the DUF and IUF designs is grounded in Expected Utility Theory and Duality Theory, ensuring that rational agents with consistent preferences should exhibit the same level of risk aversion across the two framings.

The experimental subjects comprised 88 students enrolled in various disciplines at North Carolina State University (NCSU). The experiment used a within-subjects design, where each participant completed all six treatments in randomized order to mitigate learning and order effects. In total, 88 subjects each completed 6 tasks with 10 binary choices per task, resulting in 528 tasks and 5,280 recorded decisions. The study utilized the experimental-economics software zTree, conducted on computers within the experimental-economics laboratory of the Department of Economics at NCSU. The data utilized in the present paper is derived from this experiment. On average, participants received a payoff of $16.76, inclusive of a $5 participation payment. Payoffs varied, with the lowest being $5.60 and the highest reaching $28.08. Each experimental session had an approximate duration of 75 min, with the initial 15–20 min dedicated to providing instructions. The subsequent section of the paper delves into the discussion and analysis of the study's results, particularly focusing on gender differences.

## 3. Estimation, results, and discussion: gender differences in the degrees of PaRA and PrRA

This section aims to investigate gender disparities in the degrees of PaRA and PrRA. A foundational step in this exploration involves comparing the kernel densities of the distributions of these degrees of risk aversion, which is elicited by the switching points and choice numbers selected by experimental subjects in each of the six tasks assigned. Figures 1–3 depict the histograms and kernel densities of switching points and choice numbers (which in fact represent the distributions of the degrees of risk aversion), chosen by male and female subjects under each design.

The outcomes depicted in Figure 3 reveal that, on average, women tend to exhibit higher levels of PaRA than men. This is evident across all three direct MCL designs (i.e. Tasks 1–3) in the upper row, where the red kernel density representing women's degree of PaRA lies consistently to the right of the density for men in green. This finding aligns with the stronger strand of the existing literature on the subject (e.g. Sutter *et al.*, 2013; Eckel and Grossman, 2008a, b; Dohmen *et al.*, 2011; Charness and Gneezy, 2012; Dreber *et al.*, 2011; Friedl *et al.*, 2020; Agnew *et al.*, 2008; Eckel *et al.*, 2011). Conversely, the results for the three indirect MCL designs (i.e. Tasks 4–6) in the lower row are remarkable—women generally show lower levels of PrRA than men across various MCL designs (i.e. contexts) [7], and this pattern remains robust across all the three designs. More specifically, the findings indicate that the average of the estimated midpoint CRRAs of "PaRA" is equal to *0.523 for men* (which implies "*risk-averse*" attitude, as classified by Holt and Laury (2002) and Zeytoon Nejad Moosavian *et al.* (2020)), while it is equal to *0.694 for women* (which implies "*very risk-averse*" attitude). On the contrary, the findings imply that the average of the estimated midpoint CRRAs of "PrRA" is equal to *0.755 for men* (which implies "*very risk-averse*" attitude), while it is equal to *0.646 for women* (which implies "*risk-averse*" attitude). Figure 4 visually illustrates these results.

To rigorously assess the validity and significance of the observed variations in risk aversion across genders, various statistical tests are employed in this study and reported in Appendix 2. The analysis includes the Wilcoxon-Mann-Whitney Test, Kolmogorov-Smirnov Equality-of-Distributions Test, Two-Sample *T* Test for Unpaired Data, and Ordered Probit Regression Analysis. These tests collectively provide statistical support for the notion that women tend to exhibit higher degrees of Payoff Risk Aversion (PaRA), while men generally demonstrate

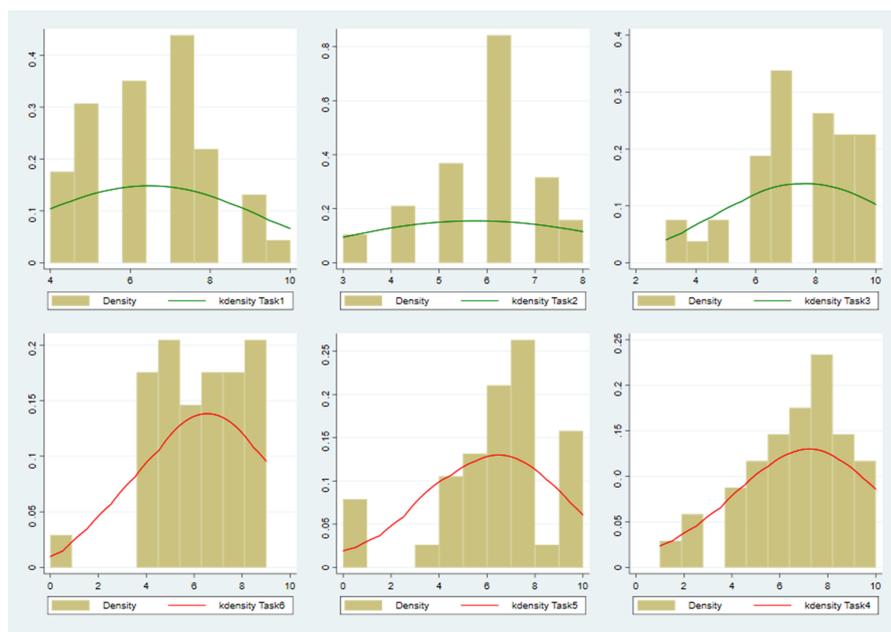

**Figure 1.** Histograms and Kernel densities of the switching points and choice numbers selected by the *male subjects*, which represent the distributions of the degrees of risk aversion exhibited by *men*. **Source:** Author's own work

higher degrees of Price Risk Aversion (PrRA). The key findings of these statistical analyses are presented in Tables A2 and A3.

The Wilcoxon-Mann-Whitney (WMW) test, also referred to as the Mann-Whitney *U* test, offers a non-parametric approach for analyzing unpaired data, serving as an alternative to the two-independent-sample *t*-test. Unlike parametric tests, the WMW test makes no assumptions about the underlying distribution of the data, only requiring that the variable is at least ordinal. Its primary objective is to assess whether there is a significant difference in the distribution of a variable between two distinct groups—in this context, women and men. Table A2 illustrates that across most direct, payoff-based designs and their average, women's levels of PaRA significantly differ from those of men. This indicates a rejection of the hypothesis that the observed gap in PaRA between women and men is due to chance alone within this experimental setting. However, concerning the degree of PrRA, while the average estimated midpoint CRRAs reveal a disparity—*0.755 for men* and *0.646 for women*—the results of the respective statistical tests on three indirect, price-based designs do not provide strong evidence to support a statistically significant difference between women's and men's degree of PrRA.

As an additional approach, we can investigate the same research question using the Kolmogorov-Smirnov (KS) equality-of-distributions test. This nonparametric statistical test assesses the equality of two distributions by testing the null hypothesis that both distributions are the same. The *p*-values associated with the KS test, outlined in Table A2, indicate that, in the case of direct, payoff-based designs, the third design and the average of all designs suggest a statistically significant difference between the degrees of PaRA exhibited by women and men. Regarding PrRA, the corresponding *p*-values suggest that across the three indirect, price-based designs, there is insufficient evidence to support a significant difference between women's and men's degrees of PrRA.

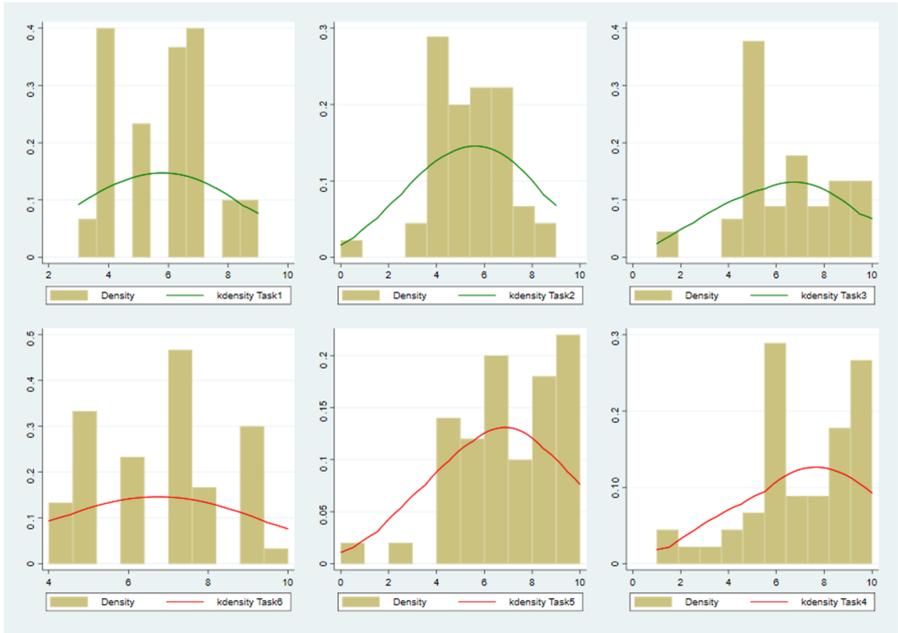

**Figure 2.** Histograms and Kernel densities of the switching points and choice numbers selected by the *female subjects*, which represent the distributions of the degrees of risk aversion exhibited by *women*. **Source:** Author's own work

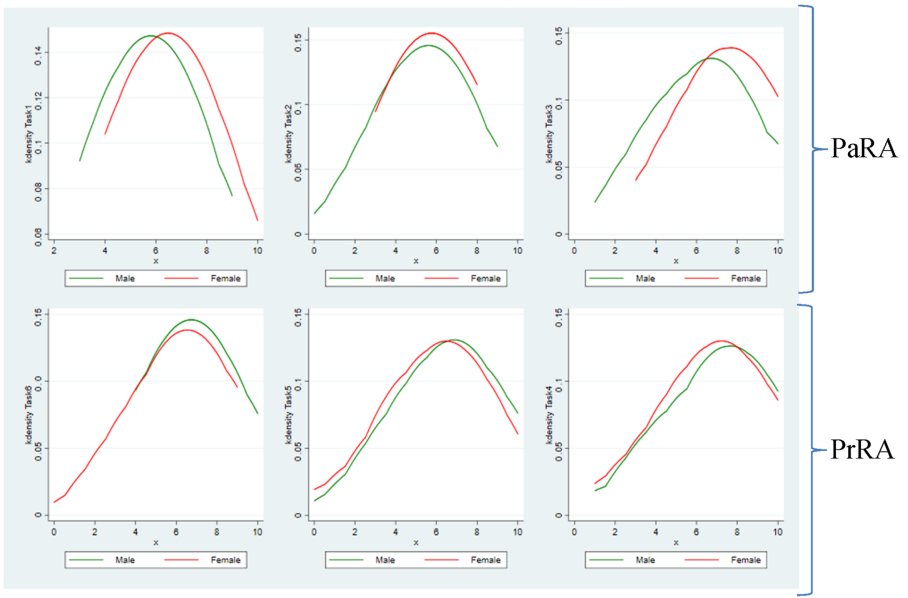

**Figure 3.** Kernel densities of the switching points and choice numbers selected by the subjects in corresponding designs, which represent the distributions of the degrees of PaRA in the upper row and the degree of PrRA in the lower row exhibited by *men* (in *green*) and exhibited by *women* (in *red*). **Source:** Author's own work

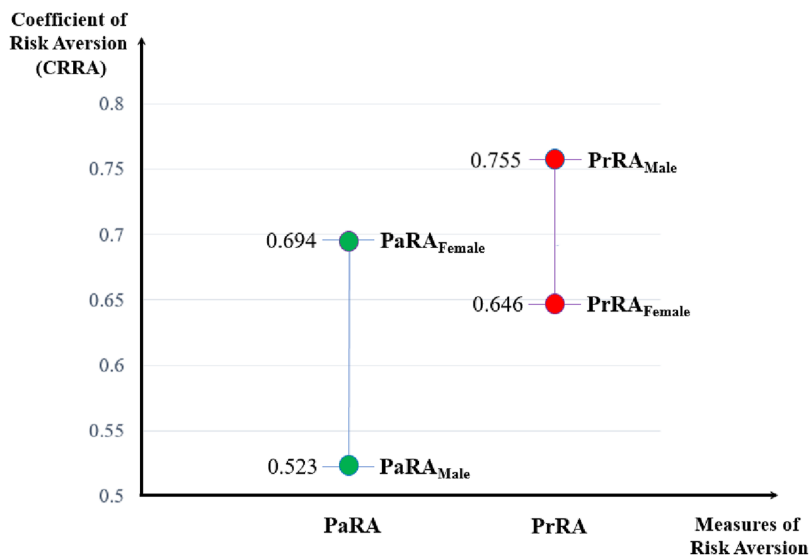

**Figure 4.** The degrees of PaRA and PrRA (measured by the coefficient of relative risk aversion or CRRA) exhibited by the experimental subjects within context across genders. **Source:** Author's own work

Alternatively, we can employ the widely used two-sample *t*-test for unpaired data. Unlike the nonparametric WMW and KS tests, the *t*-test assumes normal distribution of the variables in both groups. This parametric test, assessing the equality of means, provides an advantage in discerning which variable is statistically significantly greater. Utilizing the mid-point CRRA values of the subjects for this parametric analysis, the results align closely with those of the preceding nonparametric tests, detailed in Tables A2 and A3 in Appendix 2. These findings consistently indicate that women's degrees of PaRA are statistically significantly greater than men's, while the evidence suggesting men's PrRA being significantly greater than women's is notably weaker. In most cases, the null hypothesis can be rejected, signifying that men and women are not statistically significantly different in their risk preferences over stochastic prices, except for Task 5, in which the null hypothesis (of men being more price risk-averse than women) cannot be rejected.

Moreover, we can employ ordinal regression analysis to explore the impact of the variable gender on the degrees of PaRA and PrRA. Utilizing ordered Probit regression models, specifically designed for ordinal variables like task choice number, against the independent variable (gender), this method enables the estimation of relationships between ordinal dependent variables and one or more independent variables. Across the first three tasks (i.e. payoff-based tasks), the coefficients derived from the ordered probit model consistently show negative values, indicating that being a woman tends to elevate the degree of PaRA. This outcome is statistically significant in most payoff-based designs, including the average of the three designs. In essence, the majority of designs and their average affirm that women exhibit higher PaRA compared to men. Conversely, for the last three tasks (i.e. price-based tasks), all coefficients from the ordered Probit model are positive, indicating that being a man tends to increase the degree of PrRA. However, this result lacks statistical significance in most of the price-based designs, suggesting that men and women are not statistically significantly different in their degrees of PrRA.

In summary, the overarching observation from this section is that, on average, women tend to exhibit greater degrees of PaRA compared to men, which is a statistically significant result in most designs. Conversely, although men's PrRA appears to be higher than that of women, this outcome lacks statistical significance in numerous designs and tests, implying that men and women are not statistically significantly different in their degrees of PrRA with respect to stochastic prices. Taken together, these findings align with the third strand of the literature reviewed, positing that the degree of risk aversion of women in comparison to men may vary based on the framing and context of decision-making—e.g. Schubert *et al*. (1999), Holt and Laury (2002), and Croson and Gneezy (2009).

Now, after comparing the risk attitudes of individuals across genders within contexts (i.e. men vs. women across PaRA and PrRA, comparing one gender's PaRA with that of another gender), we can further analyze the risk attitudes of experimental subjects within each gender across contexts (i.e. comparing each gender's PaRA with their own PrRA). Figure 5 visually illustrates these results.

Figure 5 can offer two different perspectives. Firstly, the disparity between PaRA and PrRA within each gender can be viewed as an "irrationality gap," indicating the extent of irrational decision-making among experimental subjects when facing uncertainty. This is because, given Duality Theory (DT), rational individuals should exhibit equivalent degrees of PaRA and PrRA. Thus, this gap serves as a measure of irrational behavior under uncertainty, notably more pronounced in men (0.232) compared to women (0.048, which is equal to only 20% of the size of the male disparity). Secondly, this gap can be viewed as a behavioral phenomenon, indicating that men generally exhibit a greater degree of PrRA than PaRA, suggesting that they inherently have more discomfort with stochastic prices than stochastic payoffs, which can consequently mean men may display higher willingness to pay (WTP) for insurance premia guaranteeing price stability over those ensuring payoff quantities, while the reverse is true for women. The statistical significance of such an irrationality gap between PaRA and PrRA within each gender can be tested. The key findings of these statistical analyses are presented in Tables A4 and A5 in Appendix 2.

As demonstrated in Tables A4 and A5, there is a statistically significant irrationality gap between the choices that men have made under uncertainty across all designs/tasks, indicating statistically significant deviations from decisions that a rational agent would make under equivalent uncertainty about payoffs and prices. Conversely, for women, the irrationality gap

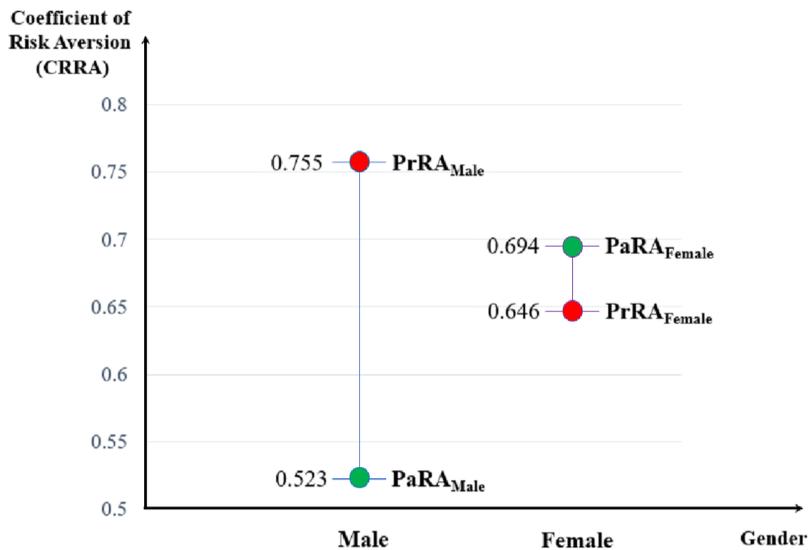

**Figure 5.** The degrees of PaRA and PrRA (measured by the coefficient of relative risk aversion or CRRA) exhibited by the experimental subjects within gender across contexts. **Source:** Author's own work

is largely non-significant across most tasks/designs and on average of the tasks. This suggests that women tend to exhibit a higher degree of rationality and consistency in their decisions under uncertainty with respect to stochastic payoffs and stochastic prices that are equivalent in essence.

The preceding results establish that, on average, men tend to exhibit a greater degree of PrRA than PaRA, while the opposite seems to hold true for women. This suggests that men tend to have a greater willingness to pay (WTP) for insurance premiums insuring price certainty over those guaranteeing payoff quantities, while the opposite holds true for women. This implies that the Risk Premium (RP), as a measure of willingness to pay for insuring uncertain situations, asked for by men is notably higher for stochastic prices compared to stochastic payoffs. In what follows, the paper delves into RP differences between men and women when faced with stochastic prices versus stochastic payoffs, terming them Price Risk Premium (PrRP) and Payoff Risk Premium (PaRP), respectively.

To contextualize RP, it is important to first understand the three notions of Expected Value (EV), Expected Utility (EU), and Certainty Equivalent (CE) within the framework of Expected Utility Theory (EUT) and lottery choice theories. RP is defined as the difference between the expected payoff of a risky situation (EV) and the certain amount for which it would be traded (CE). It represents the minimum compensation for risk acceptance and is positive for risk-averse individuals. Utilizing the CvU design, Task 2 and Task 5 allow direct elicitation and comparison of PaRP and PrRP, facilitated by the confrontation of certain and uncertain payoffs which is inherent to the CvU design.

In fact, Tasks 2 and 5 allow for the computation of CE and RP under DUF and IUF and revealing PaRP and PrRP, respectively. Further examinations of the results of these two tasks in the experiment under study indicate notable differences in risk premia between men and women. On average, men ask for a risk premium of $1.97 for Task 2 (DUF) and $3.37 for Task 5 (IUF), reflecting a considerable disparity between the two values. In contrast, women, on average, seek a risk premium of $2.27 for Task 2 (DUF) and $2.62 for Task 5 (IUF), demonstrating a close alignment between the two values. Given the true equivalency of the two tasks (despite differing contexts/framings), this consistency among women's risk-premium requests across tasks suggests a more rational and consistent approach to decision-making under uncertainty. Figure 6 demonstrates these results.

The results from this part of the study indicate that, for men, the RPs inferred from a task with uncertain prices are greater than those inferred from a task with uncertain payoffs. This is indeed another reflection of the fact that, among men, $PaRA_M < PrRA_M$. Equivalently, it can be said that, for men, CE for lotteries with payoff odds (Payoff-Certainty Equivalent, PaCE) is greater than that for lotteries with price odds (Price-Certainty Equivalent, PrCE). More formally, we can investigate the validity and significance of these findings and differences in the magnitudes of RPs using their empirical density functions and statistical hypothesis tests.

Figures 7 and 8 present histograms and kernel densities of PaRP (Task 2) and PrRP (Task 5) exhibited by male and female subjects, demonstrating the empirical assessment of RP in different risk contexts.

As shown in Figure 8, the distribution of RPs imply that men tend to ask for a greater RP when faced with random prices. This is evident by the red density line falling to the right of the green density line for men. This phenomenon is less pronounced among women, as illustrated in Figure 8. In a more formal manner, we can assess the accuracy and significance of these observations and disparities in the levels of risk premiums (RPs) through rigorous statistical hypothesis testing methods, including the Wilcoxon Signed-Rank Test, Arbuthnott-Snedecor-Cochran Sign Test, and Two-Sample $T$ Test for paired data. Tables A6 and A7 provide the main results of the above-mentioned statistical tests in an organized manner.

The first two tests are non-parametric tests in nature and the last one is a parametric test. As shown in Tables A6 and A7, all the statistical tests employed robustly support the conclusion that PrRP is statistically significantly greater than PaRP among men. Thus, it can be inferred that, for men, PaRP and PrRP are not from the same probability distribution, which strongly suggests that male individuals, in general, demand a larger premium in order to engage in lotteries with price odds compared to those with payoff odds. Consequently, the Willingness To Pay (WTP) for insurance plans covering uncertain prices exceeds that for plans covering uncertain payoffs among men. These findings underscore a departure from rational decision-making among men when confronted with random payoffs versus random prices. Hence, from a behavioral standpoint, males tend to exhibit a greater inclination towards embracing uncertain payoffs rather than uncertain prices. However, the results of the same statistical tests for women fail to demonstrate a statistically significant difference between PaRP and PrRP, suggesting a more rational approach to risk assessment among women when they are faced with inherently equivalent uncertain payoffs and uncertain prices.

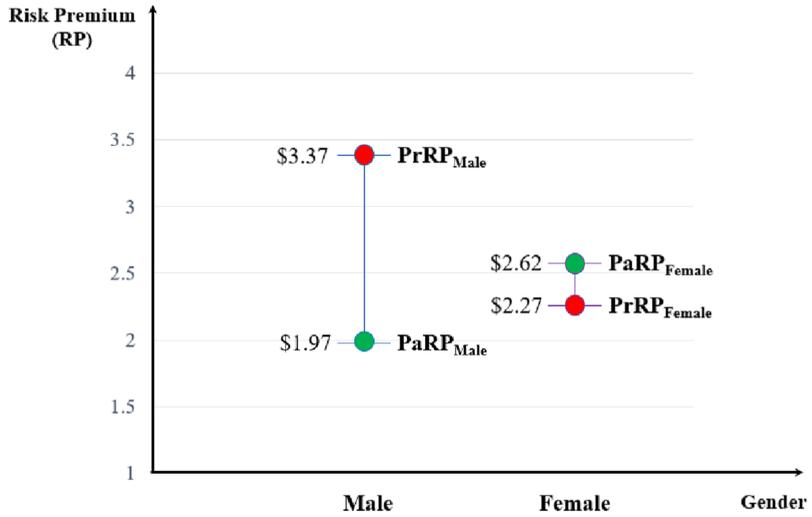

**Figure 6.** The sizes of PaRP and PrRP measured by Tasks 2 and 5 exhibited by the experimental subjects within gender across contexts. **Source:** Author's own work

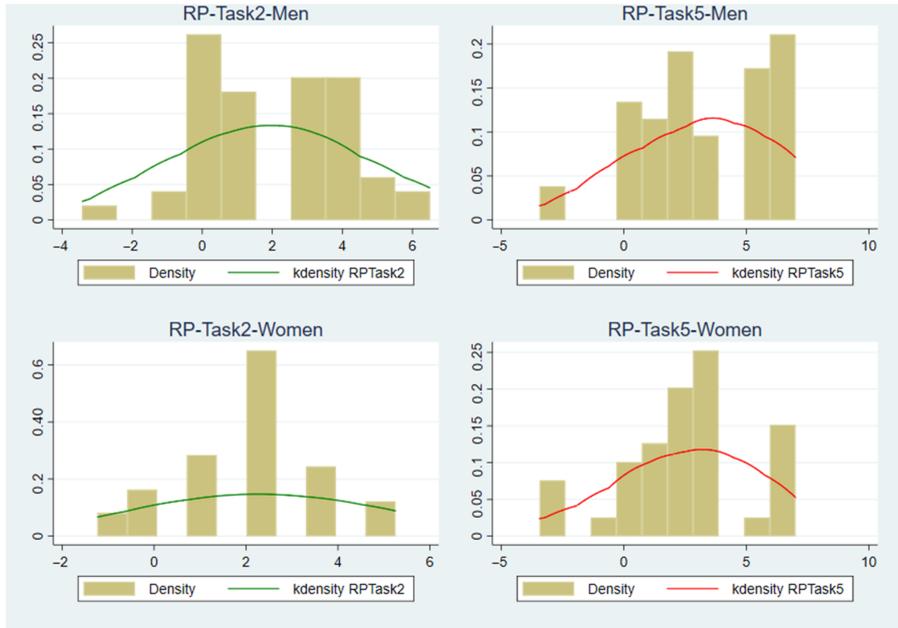

**Figure 7.** Histograms and Kernel densities of the PaRP and PrRP exhibited by the experimental subjects within gender across contexts. **Source:** Author's own work

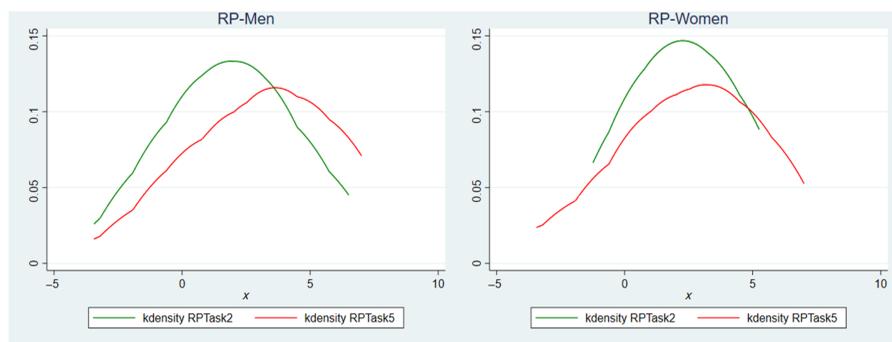

**Figure 8.** Kernel densities of the PaRP and PrRP exhibited by the experimental subjects within gender. **Source:** Author's own work

Taken altogether, the findings of this study about PaRA and PrRA across gender suggest that men tend to exhibit a greater degree of PrRA than PaRA, while the opposite seems to hold true for women.

## 4. Theoretical explanations and hypotheses for gender differences in risk aversion proposed for future research

Rather than treating behavioral explanations as post-hoc rationalizations, this section draws on established theories in behavioral economics, psychology, and gender studies to propose theoretically motivated hypotheses that may explain the observed gender differences in Price Risk Aversion (PrRA) and Payoff Risk Aversion (PaRA). While the current study does not empirically test these mechanisms, the patterns in the data align with a number of well-supported theoretical frameworks, laying the foundation for future experimental research.

*H1. The Endowment Effect and Differential Sensitivity to Ownership across Genders* – The endowment effect refers to the tendency of individuals to value an item more once they possess it (Kahneman *et al.*, 1990). In the IUF-based tasks, where participants are endowed with a monetary budget and then asked to purchase goods at uncertain prices, this sense of "ownership" may activate endowment-related loss aversion. We can hypothesize that men are more sensitive to ownership-induced loss aversion in stochastic price contexts, potentially leading to stronger PrRA. This aligns with studies showing gender asymmetries in endowment-driven preferences and loss sensitivity (e.g. Andreoni and Vesterlund, 2001; List, 2004).

*H2. Reference Dependence and Framing Asymmetries* – Prospect theory suggests that individuals evaluate outcomes relative to a reference point and weigh losses more heavily than gains (Tversky and Kahneman, 1991). In price-based tasks, the uncertain buying price may create a loss frame, while payoff-based tasks may feel more like potential gains. We can hypothesize that men's reference points in price contexts are more loss-oriented, increasing their aversion to uncertain prices. This could explain the observed irrationality gap, wherein men show greater deviation between PaRA and PrRA. Related studies show men may exhibit stronger reference dependence in market-like frames (Abeler *et al.*, 2011).

H3. *Ambiguity Aversion and Cognitive Load in Price-Based Tasks* – PrRA tasks may impose greater cognitive demands due to the indirect structure of utility under price uncertainty, making outcome computation less transparent. Ambiguity aversion— the preference for known over unknown risks—may manifest more strongly in such settings. Men may exhibit greater price-risk ambiguity aversion when the price-risk context involves greater cognitive complexity, especially in translating price uncertainty into expected value terms (Ellsberg, 1961; Fox and Tversky, 1995). This could increase PrRA among men relative to women.

H4. *Gender Differences in Socialization and Financial Risk Orientation* – Gender norms and socialization patterns have been shown to affect financial decision-making. Women are often socialized to prioritize safety, stability, and loss avoidance (Croson and Gneezy, 2009; Byrnes *et al.*, 1999). These norms may translate into higher PaRA among women, particularly in contexts involving direct gains or losses. Conversely, men may perceive greater reputational or self-evaluative costs in making suboptimal financial trades (especially under uncertainty in prices), potentially raising their PrRA.

H5. *Rationality Gap and Behavioral Consistency across Contexts* – The current study introduces the "irrationality gap"—the difference between PaRA and PrRA within individuals—as a novel behavioral marker. The data suggest this gap is larger and statistically significant for men, but not for women. It can be hypothesized that women exhibit greater consistency in their preferences across payoff and price contexts, potentially due to greater internalization of cautious decision rules or stronger reliance on heuristic-based evaluations. This hypothesis is consistent with prior findings on gender differences in decision consistency (Croson and Gneezy, 2009).

These hypotheses provide examples of opportunities for future research directions. These five explanatory hypotheses, though not tested in the current experimental setup, are firmly grounded in behavioral theory and consistent with the empirical patterns observed in the study. Each offers a potential driver of the asymmetric patterns in PaRA and PrRA across gender. Future work may design experiments that manipulate these mechanisms directly—for example, altering reference points, testing ambiguity tolerance, or controlling for endowment salience—to validate the pathways hypothesized here.

## 5. Economic and business implications of findings

If the findings of this study are replicated and validated in future research across diverse contexts, locations, demographic groups, temporal dimensions, and other pertinent factors and variables, employing diverse research approaches and methodologies to firmly establish their validity as empirically verified facts, the resulting economic and business implications of this research will be as follows.

*Gender-Specific Financial Decision Making:* Understanding these differences can help financial institutions tailor their products and services to better meet the needs and preferences of different genders. For example, investment firms can design investment options that properly match the differential risk preferences across genders.

*Risk Management Strategies:* Businesses can benefit from considering gender-specific risk attitudes when developing risk management strategies. In particular, companies may need to adopt different approaches to risk assessment and mitigation depending on whether their target market is predominantly male or female.

*Diversification of Investment Portfolios:* Investors could use these insights to diversify their portfolios effectively. For instance, women may prefer investments with more predictable returns, such as bonds or dividend-paying stocks, while men may be more willing to take on higher-risk investments in pursuit of potentially higher returns.

*Financial Education and Empowerment:* Recognizing these differences could inform financial education programs aimed at empowering individuals to make informed financial decisions. Tailored educational materials could help women overcome any hesitancy toward investing in stochastic payoffs, while encouraging men to consider the implications of stochastic prices.

*Product Innovation in Insurance Market:* Insurance companies may find opportunities to innovate and develop new financial products and services and insurance plans and policies that better cater to the risk preferences of different genders. For example, insurance companies could offer customized policies that provide varying levels of protection against stochastic payoffs or prices, allowing individuals to choose options aligned with their degrees of risk tolerance.

*Marketing Strategies:* Marketing efforts can be tailored to appeal to the risk attitudes of specific gender segments. For instance, advertising campaigns for investment products could emphasize the security and stability of returns for women, while highlighting the potential for high returns and opportunity for growth to appeal to men when it comes to the context of stochastic payoffs. They can choose to do the opposite when it comes to stochastic prices.

*Workplace Policies:* Employers can consider the impact of gender differences in risk aversion when designing compensation packages or retirement plans. Offering flexible benefits packages that accommodate different risk preferences can contribute to employee satisfaction and retention. Moreover, employers may regard women as preferable candidates for positions that entail price uncertainty and stochastic prices, given their lower level of risk aversion in such random-price contexts and their narrower irrationality gap compared to that of men.

Overall, by recognizing and accommodating gender differences in risk preferences, financial institutions and insurance companies can develop more targeted and effective financial products, investment portfolios, and insurance plans that better resonate with the preferences and priorities of their male and female customers alike. This can ultimately enhance customer satisfaction and loyalty through the utility gain that the customers will experience due to a better matching of the businesses' offerings and products with their customers' needs and preferences.

## 6. Summary and conclusion

Considerable empirical evidence in the literature of experimental economics suggests that women tend to be more risk-averse than men, often making choices associated with lower risks and lower returns. Using elicitations based on payoff-based and price-based lotteries, this study investigates gender disparities in terms of both "Payoff" Risk Aversion (PaRA) and "Price" Risk Aversion (PrRA), building upon previous research by Zeytoon Nejad Moosavian *et al.* (2020). Specifically, the findings indicate that the average midpoint CRRAs of PaRA is 0.523 for men and 0.694 for women, suggesting a "risk-averse" attitude for men and a "very risk-averse" attitude for women. Conversely, the average midpoint CRRAs of PrRA are 0.755 for men and 0.646 for women, indicating a "very risk-averse" attitude for men and just a "risk-averse" attitude for women. Additionally, the paper introduces the concept of an "irrationality gap," revealing a larger and statistically significant irrational gap for men between their induced PaRA and PrRA, while women exhibit a smaller and statistically insignificant irrationality gap. The general pattern of results from a multitude of statistical hypothesis tests confirm the significance of these gender-based differences in risk attitudes.

Additional findings reveal a significant disparity in Risk Premium (RP) across genders by computing Certainty Equivalent (CE) and Risk Premium (RP) under Direct Utility Function (DUF) and Indirect Utility Function (IUF), respectively. Men consistently exhibit higher RPs in tasks involving uncertain prices compared to uncertain payoffs, indicating a preference for certainty when prices are involved. Statistical analyses robustly support this experimental finding, highlighting a significant difference between men's RPs for uncertain prices and payoffs. Conversely, women demonstrate consistent risk-premium requests across tasks, implying a more rational and consistent approach to decision-making under uncertainty. These results underscore a departure from rationality among men but suggest a more rational risk assessment among women when faced with equivalent uncertainties in payoffs and prices.

This study sheds light on the nuanced differences in risk attitudes between men and women when faced with stochastic payoffs vs. stochastic prices. The implications of this study extend to various aspects of economic, financial, and business practices. For example, financial institutions can utilize insights into gender-specific financial decision-making to tailor their products and services accordingly in order to meet the diverse needs and preferences across genders more effectively. Businesses can benefit from considering gender-specific risk attitudes when devising risk management strategies and ensure a more effective approach that aligns with the target market's risk profiles. Investors can leverage this understanding to diversify their portfolios more effectively, catering to the differing risk preferences of men and women. Additionally, recognizing these differences can inform financial education programs and empower individuals to make more informed financial decisions. Moreover, insurance companies have opportunities for product innovation, offering customized policies aligned with gender-specific risk tolerances. Furthermore, tailored marketing strategies can appeal to the risk attitudes of specific gender segments, enhancing customer engagement. Lastly, workplace policies can be designed to accommodate gender differences in risk aversion, aligning job roles with individuals' risk attitudes to optimize performance and job satisfaction, eventually contributing to employee satisfaction and retention. Overall, recognizing and accommodating these gender differences can lead to the development of more targeted and more effective financial products and services, which in turn will enhance customer satisfaction and loyalty.

In addition to informing policy and practice, the findings of this study contribute to the broader literature on behavioral economics and decision-making under uncertainty. By uncovering gender-specific patterns in risk aversion, this research enriches our understanding of how individuals perceive and respond to risk in various contexts. The recognition of differing risk attitudes between men and women highlights the significance of integrating gender as a pivotal variable in economic and social analyses. This challenges conventional assumptions of uniform risk behavior across demographic groups and rejects the notion that women are universally more risk-averse than men across all contexts. Moreover, the implications of gender differences in risk attitudes can extend beyond applied domains and social awareness to broader societal dynamics. Recognizing and addressing disparities in risk perception and decision-making can pave the way for more inclusive policies and systems. By promoting awareness and understanding of gender-specific risk preferences, policymakers can work towards creating environments that foster equal opportunities for individuals of all genders having different risk attitudes, which can ultimately contribute to societal well-being and progress.

Moving forward, future research should delve deeper into the underlying psychological mechanisms driving gender differences in risk aversion. By examining factors such as cognitive biases, social norms, and individual experiences, researchers can develop a more comprehensive understanding of how gender influences risk preferences. Overall, this study highlights the significance of interdisciplinary research approaches that integrate insights from psychology, economics, and gender studies. By synthesizing empirical evidence and theoretical frameworks from diverse fields, researchers can develop more nuanced and contextually relevant understandings of human behavior under uncertainty.

**Data availability statement**
The data that support the findings of this study are available from the corresponding author upon reasonable request.

**Appendix 1**
**Literature review**
This literature review unfolds into two sections. The first section undertakes a concise review of the experimental economics literature, shedding light on key methodologies and experimental designs employed for studying risk attitudes in past studies. This exploration serves as a contextual backdrop for the methodology used in this paper. The second section delves into the findings of related papers in the literature, specifically focusing on gender differences in risk aversion and categorizing these findings into distinct groups to distill valuable insights into the differential risk preferences between men and women. This comprehensive overview lays the groundwork for identifying gaps within the current literature and emphasizing the importance of addressing the research objectives of the present study.

The exploration of risk aversion within the realm of experimental economics has been enriched by the deliberate deployment of experimental methodologies and designs in past studies. These meticulously crafted approaches not only have provided nuanced insights into the intricate aspects of risk attitudes but have also played a pivotal role in shaping the landscape of experimental research within this domain. In what follows, three popular MCL methods that are extensively used to understand risk aversion are explained [8].

In their influential work, Holt and Laury (2002) introduced a pivotal experimental design to assess individuals' risk attitudes, which later shaped many subsequent experiments in the field. The study addressed a gap in understanding how risk aversion should be modeled within standard economic theories. Utilizing a menu of choices, they gauged risk aversion and explored behavior under real and hypothetical incentives for lotteries of varying magnitudes. Their results provided substantial evidence for risk aversion even at lower payoff levels, challenging assumptions of risk neutrality in economic agents' behavior when payoff levels are low. Similarly, Binswanger (1980) pioneered the use of an MCL experimental design to study farmers' risk attitudes in rural India. The findings of his study highlighted widespread risk aversion among farmers. He showed that the degree of risk aversion increased as payoffs escalated, and challenged assumptions about the rarity of extreme risk aversion. Balsa *et al.* (2015) contributed to the literature by introducing a Certainty-versus-Uncertainty (CvU) design. Their experiment, akin to Holt and Laury's MPL design with some modifications to provide a certain option vs. an uncertain option in each decision, investigated whether students' risk attitudes were influenced by peers, revealing a substantial impact, particularly among male adolescents. Their evidence pointed to strong peer effects in risk aversion, which underscores the importance of considering social influences when studying individual risk attitudes. The present paper employs the three aforementioned methods in its analysis. These methods not only form a proper foundation for comprehensively examining risk attitudes but also help in checking for robustness of the results with respect to contextual structures. That is, these methods enable us to navigate and interpret the complexities inherent in risk aversion analysis under the Expected Utility Theory (EUT), as these three methods that were introduced above can address "probability weighting", "varying payoffs", and "certainty vs. uncertainty weighing", respectively, and these are three important facets of EUT.

Gender differences in risk aversion have been a subject of extensive research in the past few decades, mainly aiming to unravel the intricacies of how men and women approach uncertainty and make decisions under risk. A multitude of studies have delved into this intriguing realm, seeking to detect whether a significant discrepancy exists in the degree of risk aversion between genders. The remainder of this literature review categorizes such findings into three distinct groups: (1) studies suggesting that women tend to be more risk-averse, (2) those indicating no statistically significant gender differences, and (3) those suggesting that the degree of risk aversion of women in comparison to men may vary based on the framing and context of decision-making.

Multiple studies support the claim that women generally exhibit higher levels of risk aversion compared to men. For example, Dohmen *et al.* (2011) investigate the relationship between individual risk attitudes, gender, and other demographics in different contexts, using a large representative survey and a complementary experiment through paid lottery choices, finding that women are less willing to take risks than men. Sutter *et al.* (2013) study risk attitudes, ambiguity attitudes, and time preferences of adolescents using experiments with standard choice list tasks, finding that girls are more risk-averse than boys. Eckel and Grossman (2008a, b) use experiments on risk aversion in the context of abstract gambling experiments, contextual environment experiments, and field experiments to find evidence of systematic gender differences. They explore risk aversion through a review of experiments, and find that, in most studies, women are found to be more risk-averse than men.

A second set of studies suggests that there may be no statistically significant difference in the degree of risk aversion between men and women. For example, Sarin and Wieland (2016) investigate peoples' subjective probabilities and gender differences in risk aversion using five different experiments including games of chance and Becker-DeGroot-Marschak (BDM) procedures. They find that men and women value bets similarly both before and after controlling for subjective probabilities, so women are not more risk-averse than men for bets on real events. Harrison *et al.* (2007) study individual risk attitudes in Denmark through controlled experiments using a representative sample and considering a multitude of socio-demographic characteristics. By exploring a multitude of socio-demographic characteristics in identifying gender differences in risk aversion, they find that men and women are not statistically significantly different in their degrees of risk aversion. Hillesland (2019) studies gender differences in risk preferences through an analysis of asset allocation decisions in the context of a developing country, finding that men and women are not statistically significantly different in their degrees of risk aversion.

In contrast, a third set of studies posits that the degree of risk aversion of women in comparison to men may vary based on the framing and context of decision-making. Croson and Gneezy (2009) delve into gender differences in risk preferences to give insights into the influence of gender on economic decision-making, and uncover that, on average, women tend to be more risk-averse than men in general. However, they assert that there are exceptions to this pattern, particularly in the case of professional men and women, whose risk attitudes do not typically exhibit significant differences. Holt and Laury (2002) examine risk aversion, incentive effects, and payoff scale effect using a menu of paired lottery choices, finding that men are slightly less risk-averse, but this gender effect fades away when payoffs become larger. That is, females exhibit a stronger level of risk aversion than males in low-payoff settings, but their differences are insignificant in high-payoff settings. Schubert *et al.* (1999) study financial decision-making under risk and gender differences in risk aversion using an experiment with gambling decisions, finding that women do not generally make less risky financial choices than men. Instead, they show that women are more risk-averse than men in the gain-domain frame, but men are more risk-averse than women in the loss-domain gambles. This set of studies indicate that the degree of risk aversion in different genders can depend heavily on the decision frames and contexts.

Table A1 summarizes the results from these three groups of studies and adds information about numerous additional studies conducted in this area.

As reported in Table A1, a significant body of research consistently suggests that women exhibit higher levels of risk aversion compared to men. These studies, employing a diverse range of methodologies such as experiments and surveys, have collectively illuminated a pattern wherein women consistently display a higher degree of risk aversion with respect to financial risks. Whether exploring varied demographics or experimenting with different innovative research designs, the key findings consistently point towards a gender disparity where women tend to be less tolerant of uncertainty in financial decision-making.

On the other hand, another group of studies emphasizes that gender differences in risk aversion may not be as pronounced. These studies, employing various approaches such as lab experiments, field experiments, and surveys examining different risk scenarios, collectively find that the differences observed in risk attitudes between men and women are either not statistically significant or are context-dependent. The implications drawn from these findings suggest that gender may not be a decisive factor in shaping individual risk attitudes, challenging the notion of a uniform gender-based predisposition towards risk aversion. These studies contribute to a nuanced understanding of the complex interplay of factors influencing risk preferences, where gender may play a less deterministic role in determining individual risk-taking behavior.

To conclude, after a thorough examination of the literature on risk attitude elicitation, a notable observation emerges—there is an absence of studies exploring gender differences in "price" risk aversion (PrRA). As introduced by Zeytoon Nejad Moosavian *et al.* (2020), the context of the Indirect Utility Function (IUF) can be used to achieve this end. Therefore, the unresolved question in this body of literature pertains to whether women are also more price risk-averse than men or not. This identified gap serves as the focal point that the current paper aims to address, leveraging data collected and the methodology introduced by Zeytoon Nejad Moosavian *et al.* (2020).

**Table A1.** An overview of the existing literature on the topic of gender differences in risk aversion

| Study | Authors | Year | Focus | Methodology | Key findings | Contributions to or implications for gender differences | Conclusion on gender disparities in the degree of risk aversion |
|---|---|---|---|---|---|---|---|
| 1. Impatience and uncertainty: experimental decisions predict adolescents' field behavior | Sutter *et al*. | 2013 | Risk attitudes, ambiguity attitudes, and time preferences of adolescents | Experiments with standard choice list tasks | Impatience is found to be a strong predictor of health-related field behavior and saving decisions, and that girls are more risk averse than boys | Offers new insights into the relationship between risk attitudes, ambiguity attitudes, and time preferences (i.e. impatience) | Females > Males |
| 2. Men, women and risk aversion: experimental evidence | Eckel and Grossman | 2008a, b | Experiments on risk aversion in the context of abstract gamble experiments, contextual environment experiments, and field experiments to find evidence of systematic gender differences | A review of the results from experimental measures of risk aversion | Explores risk aversion through a review of experiments, providing evidence of systematic differences in the behavior of men and women. In most studies, women are found to be more averse to risk than men | Provides invaluable insights into how men and women perceive and approach risky decisions | Females > Males |
| 3. Individual risk attitudes: measurement, determinants, and behavioral consequences | Dohmen *et al*. | 2011 | Relationship between individual risk attitudes, gender, and other demographics in different contexts | A large representative survey and a complementary experiment using paid lottery choices | Finds that gender, age, height, and parental background have economically significant impacts on willingness to take risks. Women are less willing to take risks than men, so women are more risk-averse | Gives insights into the diversity of risk preferences within populations, with attention to gender-specific patterns, and finds similar results on the determinants of risk attitudes in different contexts | Females > Males |

**Table A1.** Continued

| Study | Authors | Year | Focus | Methodology | Key findings | Contributions to or implications for gender differences | Conclusion on gender disparities in the degree of risk aversion |
|---|---|---|---|---|---|---|---|
| 4. Strong evidence for gender differences in risk taking | Charness and Gneezy | 2012 | Gender differences in risk-taking in the context of financial decision making by using data assembled from 15 sets of experiments | Experiments with a simple underlying investment game | Presents robust findings on gender disparities in risk-taking behaviors, finding that women invest less, and therefore seem to be more financially risk-averse than men | Contributes empirical evidence to the discourse on gender differences in risk-taking behaviors in the context of financial decisions | Females > Males |
| 5. Dopamine and risk choices in different domains: findings among serious tournament bridge players | Dreber *et al.* | 2011 | Risk taking in the card game contract bridge, and economic risk taking as proxied by a financial gamble | Experiments including a card game and a financial gamble | Finds evidence that men take more overall risk in bridge than women, and strong evidence that men take more economic risk in an investment game. It also finds that the dopamine system plays an important role in explaining individual differences in risk taking | Identifies a strong interaction among desirable risk-taking behavior, measured success, and genetic variation | Females > Males |
| 6. Gender differences in social risk taking | Friedl *et al.* | 2020 | Risk taking in social contexts | Controlled experiments | Finds that inequality aversion is the main driver for risk aversion in social risk taking. This effect is mainly driven by strong inequality aversion of women | Concludes that gender differences in social risk taking are culture-specific, and that women are more risk-averse than men when payoffs are unequally distributed | Females > Males |

(*continued*)

| Study | Authors | Year | Focus | Methodology | Key findings | Contributions to or implications for gender differences | Conclusion on gender disparities in the degree of risk aversion |
|---|---|---|---|---|---|---|---|
| 7. Who chooses annuities? An experimental investigation of the role of gender, framing, and defaults | Agnew *et al.* | 2008 | The role of gender, framing and defaults in risk attitudes among people facing financial decision about retirement plans | A controlled experiment | Discovers that women are more likely to choose annuities which is partly explained by differences in risk aversion and financial literacy | Contributes to the literature by studying the role of gender in important financial decisions such as retirement, and the choice between purchasing an annuity (a safe option) or investing savings (a risky option) on their own | Females > Males |
| 8. On the development of risk preferences: experimental evidence | Eckel *et al.* | 2011 | A field experiment eliciting the risk preferences among adolescents and examining various factors influencing the development of these risk preferences | A field experiment | Finds that girls are more risk-averse than boys, and also finds some evidence for a peer effect and a school-quality effect | Confirms that factors previously considered by economists (e.g. gender, ethnicity, height, and parental education) are found to be stronger determinants of risk preferences than those proposed by cognitive development theory and emotional development theory | Females > Males |

**Table A1.** Continued

| Study | Authors | Year | Focus | Methodology | Key findings | Contributions to or implications for gender differences | Conclusion on gender disparities in the degree of risk aversion |
|---|---|---|---|---|---|---|---|
| 9. Gender differences in risk behavior in financial decision-making: an experimental analysis | Powell and Ansic | 1997 | Sensitivity of differences in risk preference to the framing of tasks and level of task familiarity to subjects | Computerized laboratory experiments | By examining gender differences in risk propensity and strategy in financial decision-making, it shows that women are less risk-seeking than men regardless of familiarity with tasks, task framing, costs, or ambiguity | Shows that the framing of tasks is not important in and that gender differences may arise in different framings of ambiguity. Women are more ambiguity-averse than men in the context of investment, but not in the context of insurance | Females > Males (in the context of investment) Females ~ Males (in the context of insurance) |
| 10. Risk Aversion and Incentive Effects | Holt and Laury | 2002 | Risk aversion, incentive effects, payoff scale effect, introducing a menu of paired lottery choices, contrasting hypothetical and real payoffs, and fitting a hybrid power utility function with increasing relative and decreasing absolute risk aversion | An experiment including a menu of paired lottery choices | Finds evidence that, with usual low laboratory payoffs of several dollars, most subjects are risk-averse, and that scaling up all payoffs by factors of twenty to ninety makes little difference when the high payoffs are hypothetical, but a big difference when the payoffs are real, where subjects become more risk-averse | Creates a menu of paired lottery choices which is structured so that the switching point from risky choices to safe choices can be used to infer the degree of risk aversion. It also shows that men are slightly less risk-averse, but this gender effect fades away when payoffs become larger | Females > Males (in low-payoff settings) Females ~ Males (in high-payoff settings) |

(*continued*)

**Table A1.** Continued

| Study | Authors | Year | Focus | Methodology | Key findings | Contributions to or implications for gender differences | Conclusion on gender disparities in the degree of risk aversion |
|---|---|---|---|---|---|---|---|
| 11. Financial Decision-Making: Are Women Really More Risk Averse? | Schubert *et al.* | 1999 | Financial decision-making under risk and gender differences in risk aversion and comparative risk propensity of male and female subjects in financial choices | An experiment with gambling decisions, and financially motivated risky decisions embedded in an investment or insurance context | Challenges stereotypes about women being more risk-averse | Finds that women do not generally make less risky financial choices than men. It also shows that the comparative risk propensity of male and female subjects in financial choices strongly depends on the decision frames and contexts, and that women were more risk-averse than men in the gain-domain frame, but the opposite holds true for the loss-domain gambles | Females > Males (in gain-domain gambles) Females < Males (in loss-domain gambles) |
| 12. Gender Differences in Preferences | Croson and Gneezy | 2009 | Gender differences in preferences (Risk taking, social preferences, and reaction to competition) | A survey of a series of economics experiments | Uncovers distinct inclinations and risk attitudes among men and women | Contributes to our understanding of how gender influences economic decision-making in three areas | Females > Males (in general) Females ~ Males (among professionals) |

**Table A1.** Continued

| Study | Authors | Year | Focus | Methodology | Key findings | Contributions to or implications for gender differences | Conclusion on gender disparities in the degree of risk aversion |
|---|---|---|---|---|---|---|---|
| 13. Does Knowledge of Finance Mitigate the Gender Difference in Financial Risk-Aversion? | Hibbert *et al.* | 2013 | The impact of financial knowledge on gender differences in financial risk-aversion by comparing finance professors' portfolio allocations to those of respondents in the Federal Reserve's Survey of Consumer Finances (SCF) | A survey of finance professors from US universities | Showing that, among highly educated individuals, women display higher levels of risk aversion, but with a high level of financial education, the gender difference in financial risk aversion diminishes, leading to equal likelihood in investing in risky assets for both men and women | Provides evidence on how general education and financial education influence the gender difference in financial risk aversion | Females > Males (among the highly educated) Females ~ Males (among those with a high level of financial education) |
| 14. Gender Specific Attitudes towards Risk and Ambiguity: An Experimental Investigation | Schubert *et al.* | 2000 | Financial decision-making under ambiguity in probability information sets and gender differences in ambiguity aversion within investment and insurance contexts | A set of lottery experiments with three types of probability information in an investment or insurance context | Shows that the framing of information is important and that gender differences may arise in different framings of ambiguity. Women are more ambiguity-averse than men in the context of investment, but not in the context of insurance | Examines gender differences in ambiguity aversion, showing that women do not generally choose to make safer financial decisions than men, and that there are different forms of gender differences in different decision frames and contexts | Females ~ Males (in terms of risk aversion) Females > Males (ambiguity aversion in the investment context) Females < Males (ambiguity aversion in the insurance context) |

(*continued*)

**Table A1.** Continued

| Study | Authors | Year | Focus | Methodology | Key findings | Contributions to or implications for gender differences | Conclusion on gender disparities in the degree of risk aversion |
|---|---|---|---|---|---|---|---|
| 15. Gender and Risk: Women, Risk Taking and Risk Aversion | Maxfield *et al.* | 2010 | Exploring women's risk taking and reasons for stereotype persistence | Survey (Uses the Simmons Gender and Risk Survey database of 661 female managers) | Finds evidence of gender neutrality in risk propensity and decision-making in several managerial contexts other than portfolio allocation | Provides insights into how context can influence gender differences in risk taking and synthesizes evidence on risk taking and gender in different contexts | Females ∼ Males (in all contexts but portfolio allocation) Females > Males (in the context of portfolio allocation) |
| 16. Gender Effects for Loss Aversion: Yes, No, Maybe? | Bouchouicha *et al.* | 2019 | Gender differences in loss aversion in terms of four different definitions of loss aversion | Used data from Vieider *et al.* (2016), which, in turn, used controlled experiments in 30 countries to estimate certainty equivalents using lotteries | Indicates that across various definitions of loss aversion, males tend to exhibit greater aversion to losses than females in most cases. In other definitions, either females display higher levels of loss aversion, or the observed difference is statistically insignificant | Employs four definitions of loss aversion in the literature to investigate gender effects in loss aversion, showing that gender differences are sensitive to the definition of loss aversion used in an experiment | Females < Males (in case of 2 definitions of loss aversion) Females > Males (in terms of 1 definition of loss aversion) Females ∼ Males (in terms of 1 definition of loss aversion) |

**Table A1.** Continued

| Study | Authors | Year | Focus | Methodology | Key findings | Contributions to or implications for gender differences | Conclusion on gender disparities in the degree of risk aversion |
|---|---|---|---|---|---|---|---|
| 17. Risk aversion for Decisions under Uncertainty: Are There Gender Differences? | Sarin and Wieland | 2016 | Peoples' subjective probabilities and gender differences in risk aversion | Five different experiments including games of chance and BDM procedures | Finds that men and women value bets similarly both before and after controlling for subjective probabilities, so women are not more risk-averse than men for bets on real events | Introduces a method that controls for subjective probability in comparing gender differences in risk aversion. It also finds that subjective probability is the main driver of valuation for people when making decisions under uncertainty | Females ∼ Males |
| 18. Estimating Risk Attitudes in Denmark: A Field Experiment | Harrison *et al.* | 2007 | Studying individual risk attitudes in Denmark through controlled experiments using a representative sample and considering a multitude of socio-demographic characteristics | Controlled experiments in the field | Exploration of a multitude of socio-demographic characteristics in identifying gender differences in risk aversion, finding that men and women are not statistically significantly different in their degrees of risk aversion | Finds that the average Dane is risk-averse, and that risk attitudes vary significantly with respect to several important socio-demographic variables | Females ∼ Males |

**Table A1.** Continued

| Study | Authors | Year | Focus | Methodology | Key findings | Contributions to or implications for gender differences | Conclusion on gender disparities in the degree of risk aversion |
|-------|---------|------|-------|-------------|--------------|--------------------------------------------------------|----------------------------------------------------------------|
| 19. Attitudes toward Risk: Theoretical Implications of an Experiment in Rural India | Binswanger | 1981 | Risk aversion among farmers in India | Large-Scale field experiments with MCL-like experimental designs | Finds that most farmers exhibit a considerable degree of risk aversion that has a tendency to increase as payoffs are scaled up, and that most individuals have very similar levels of risk aversion, and that wealth has trivial effect on the degree of risk aversion and years of schooling reduce the extent of risk aversion | Creates one of the first MCL experimental designs ever to infer risk attitudes and check a multitude of demographics such as gender, wealth, and years of schooling | Females ∼ Males |
| 20. Gender Differences in Risk Behavior: An Analysis of Asset Allocation Decisions in Ghana | Hillesland | 2019 | Gender differences in risk preferences through an analysis of asset allocation decisions in the context of a developing country | National surveys and a decomposition method | Contributes to the understanding of national-representative sex-disaggregated data with self-reported asset ownership and wealth information of individuals, finding that men and women are not statistically significantly different in their degrees of risk aversion | Provides insights into gender differences in risk preferences within a developing country | Females ∼ Males |

**Note(s):** "Females > Males" means women are found to be more risk-averse than men. "Females < Males" means the opposite. And finally, "Females ∼ Males" means the difference between their degrees of risk aversion is found to be statistically insignificant
**Source(s):** Author's own work

Review of Behavioral Finance

**Table A2.** Summary of the results of the statistical tests regarding the distinct degrees of *PaRA* across genders

| Test | Test explanation | Test hypothesis | Task1 (PaRA) (Women = Men) | Task2 (PaRA) (Women = Men) | Task3 (PaRA) (Women = Men) | Women = Men (average of Pa-Tasks) | Overall conclusion |
|---|---|---|---|---|---|---|---|
| Wilcoxon-Mann-Whitney test | It tests whether the distributions in two groups are the same (non-parametric) | $H_0$: Both distributions are the same | Reject $H_0$ at 10% Prob>\|z\| = 0.0596 $H_1$: Women≠Men Confirm $H_1$ | Fail to Reject $H_0$ Prob>\|z\| = 0.4128 $H_1$: Women≠Men Cannot Confirm $H_1$ | Reject $H_0$ at 10% Prob>\|z\| = 0.0299 $H_1$: Women≠Men Confirm $H_1$ | Reject $H_0$ at 10% Prob>\|z\| = 0.0687 $H_1$: Women≠Men Confirm $H_1$ | It shows that most designs as well as their average confirm that *Women's PaRA ≠ Men's PaRA* |
| Kolmogorov-Smirnov equality-of-distributions test | It tests the equality of distributions (non-parametric) | $H_0$: Both distributions are the same | Fail to Reject $H_0$ Co. *P*-value = 0.525 $H_1$: Women≠Men Cannot Confirm $H_1$ | Fail to Reject $H_0$ *P*-value = 0.622 $H_1$: Women≠Men Cannot Confirm $H_1$ | Reject $H_0$ at 10% *P*-value = 0.033 $H_1$: Women≠Men Confirm $H_1$ | Reject $H_0$ at 10% *P*-value = 0.073 $H_1$: Women≠Men Confirm $H_1$ | It shows that the 3rd design as well as the average of the designs confirm that *Women's PaRA ≠ Men's PaRA* |
| Two-sample t-test for unpaired data (using mid-point CRRA's) | It tests the equality of the means of a normally-distributed variable for two independent groups (parametric) | $H_0$: The mean of the difference is zero | Reject $H_0$ at 10% Prob $(T > t) = 0.0249$ $H_1$: Women > Men Confirm $H_1$ | Fail to Reject $H_0$ Prob $(T > t) = 0.2247$ $H_1$: Women > Men Cannot Confirm $H_1$ | Reject $H_0$ at 10% Prob $(T > t) = 0.0139$ $H_1$: Women > Men Confirm $H_1$ | Reject $H_0$ at 10% Prob $(T > t) = 0.0189$ $H_1$: Women > Men Confirm $H_1$ | It shows that most designs as well as their average confirm that *Women's PaRA > Men's PaRA* |

(*continued*)

**Table A2.** Continued

| Test | Test explanation | Test hypothesis | Task1 (PaRA) (Women = Men) | Task2 (PaRA) (Women = Men) | Task3 (PaRA) (Women = Men) | Women = Men (average of Pa-Tasks) | Overall conclusion |
|---|---|---|---|---|---|---|---|
| Regression analysis using ordered Probit model | It estimates relationships between an ordinal dependent variable and one or more independent variable(s) (An ordinal regression analysis) | $H_0$: The difference between men's degree of PaRA and that of women is zero | Reject $H_0$ at 10% Prob>\|z\| = 0.035 *Coef* = −0.470 S.E. = 0.223 $Z$ = −2.10 95% C.I. = [−0.908, −0.032] $H_1$: Women > Men Confirm $H_1$ | Fail to Reject $H_0$ Prob>\|z\| = 0.527 *Coef* = −0.140 S.E. = 0.222 $Z$ = −0.63 95% C.I. = [−0.574, 0.294] $H_1$: Women > Men Cannot Confirm $H_1$ | Reject $H_0$ at 10% Prob>\|z\| = 0.044 *Coef* = −0.452 S.E. = 0.224 $Z$ = −2.02 95% C.I. = [−0.891, −0.128] $H_1$: Women > Men Confirm $H_1$ | Reject $H_0$ at 10% Prob>\|z\| = 0.096 *Coef* = −0.365 S.E. = 0.219 $Z$ = −1.67 95% C.I. = [−0.796, 0.065] $H_1$: Women > Men Confirm $H_1$ | It shows that most designs as well as their average confirm that *Women's PaRA > Men's PaRA* |

**Source(s):** Author's own work

**Table A3.** Summary of the results of the statistical tests. Regarding the distinct degrees of *PrRA* across genders

| Test | Test explanation | Test hypothesis | Task4 (PrRA) (Women = Men) | Task5 (PrRA) (Women = Men) | Task6 (PrRA) (Women = Men) | Women = Men (average of Pr-Tasks) | Overall conclusion |
|------|------------------|-----------------|----------------------------|----------------------------|----------------------------|-----------------------------------|--------------------|
| Wilcoxon-Mann-Whitney test | It tests whether the distributions in two groups are the same (non-parametric) | $H_0$: Both distributions are the same | Fail to Reject $H_0$ Prob>|z| = 0.3597 $H_1$: Women≠Men Cannot Confirm $H_1$ | Fail to Reject $H_0$ Prob>|z| = 0.2703 $H_1$: Women≠Men Cannot Confirm $H_1$ | Fail to Reject $H_0$ Prob>|z| = 0.4776 $H_1$: Women≠Men Cannot Confirm $H_1$ | Fail to Reject $H_0$ Prob>|z| = 0.1978 $H_1$: Women≠Men Cannot Confirm $H_1$ | It shows that under these three designs there is NOT strong evidence to suggest that Women's PrRA ≠ Men's PrRA |
| Kolmogorov-Smirnov equality-of-distributions test | It tests the equality of distributions (non-parametric) | $H_0$: Both distributions are the same | Fail to Reject $H_0$ *P*-value = 0.613 $H_1$: Women≠Men Cannot Confirm $H_1$ | Fail to Reject $H_0$ *P*-value = 0.267 $H_1$: Women≠Men Cannot Confirm $H_1$ | Fail to Reject $H_0$ *P*-value = 0.973 $H_1$: Women≠Men Cannot Confirm $H_1$ | Fail to Reject $H_0$ *P*-value = 0.776 $H_1$: Women≠Men Cannot Confirm $H_1$ | It shows that under these three designs there is NOT much evidence to suggest that Women's PrRA ≠ Men's PrRA |
| Two-sample t-test for unpaired data (using mid-point CRRA's) | It tests the equality of the means of a normally-distributed variable for two independent groups (parametric) | $H_0$: The mean of the difference is zero | Fail to Reject $H_0$ Prob (T < *t*) = 0.3098 $H_1$: Women≠Men Cannot Confirm $H_1$ | Reject $H_0$ at 10% Prob (T < *t*) = 0.0932 $H_1$: Women≠Men Confirm $H_1$ | Fail to Reject $H_0$ Prob (T < *t*) = 0.2039 $H_1$: Women≠Men Cannot Confirm $H_1$ | Fail to Reject $H_0$ Prob (T < *t*) = 0.1315 $H_1$: Women≠Men Cannot Confirm $H_1$ | It shows that under these three designs (except for Task 5) there is NOT strong evidence to suggest that Women's PrRA < Men's PrRA |
| Regression analysis using ordered Probit model | It estimates relationships between an ordinal dependent variable and one or more independent variable(s) (An ordinal regression analysis) | $H_0$: The difference between men's degree of PrRA and that of women is zero | Fail to Reject $H_0$ Prob>|z| = 0.319 *Coef = 0.221* S.E. = 0.222 Z = 1.00 95% C.I. = [−0.214, 0.655] $H_1$: Women≠Men Cannot Confirm $H_1$ | Fail to Reject $H_0$ Prob>|z| = 0.225 *Coef = 0.267* S.E. = 0.221 Z = 1.21 95% C.I. = [−0.165, 0.700] $H_1$: Women≠Men Cannot Confirm $H_1$ | Fail to Reject $H_0$ Prob>|z| = 0.326 *Coef = 0.218* S.E. = 0.222 Z = 0.98 95% C.I. = [−0.218, 0.654] $H_1$: Women≠Men Cannot Confirm $H_1$ | Fail to Reject $H_0$ Prob>|z| = 0.150 *Coef = 0.313* S.E. = 0.218 Z = 1.44 95% C.I. = [−0.114, 0.739] $H_1$: Women≠Men Cannot Confirm $H_1$ | It shows that under these three designs, there is NOT strong evidence to suggest that Women's PrRA < Men's PrRA |

**Table A4.** Comparative analysis of PaRA and PrRA across contexts: key findings from three statistical tests for *Men*

| Test | Test explanation | Test hypothesis | Men Task1 = Task6 (PaRA=PrRA) | Men Task2 = Task5 (PaRA=PrRA) | Men Task3 = Task4 (PaRA=PrRA) | Men PaRA=PrRA (average of Pa- and Pr-Tasks) | Overall conclusion |
|---|---|---|---|---|---|---|---|
| Wilcoxon matched-pairs signed-ranks test | It tests the equality of matched pairs of observations (non-parametric) | $H_0$: Both distributions are the same | Reject $H_0$ at 10% Prob>$|z|$ = 0.0056 $H_1$: $PaRA_M \neq PrRA_M$ Confirm $H_1$ | Reject $H_0$ at 10% Prob>$|z|$ = 0.0020 $H_1$: $PaRA_M \neq PrRA_M$ Confirm $H_1$ | Reject $H_0$ at 10% Prob>$|z|$ = 0.0355 $H_1$: $PaRA_M \neq PrRA_M$ Confirm $H_1$ | Reject $H_0$ at 10% Prob>$|z|$ = 0.0005 $H_1$: $PaRA_M \neq PrRA_M$ Confirm $H_1$ | It shows that all of the designs as well as their average confirm that $PaRA_M \neq PrRA_M$ |
| Arbuthnott-Snedecor-Cochran sign test | It tests the equality of matched pairs of observations (non-parametric) | $H_0$: The median of the differences is zero (the true proportion of positive (negative) signs is one-half) | Reject $H_0$ at 10% Prob(.) = 0.0243 $H_1$: $PaRA_M \neq PrRA_M$ Confirm $H_1$ | Reject $H_0$ at 10% Prob(.) = 0.0336 $H_1$: $PaRA_M \neq PrRA_M$ Confirm $H_1$ | Reject $H_0$ at 10% Prob(.) = 0.0652 $H_1$: $PaRA_M \neq PrRA_M$ Confirm $H_1$ | Reject $H_0$ at 10% Prob(.) = 0.0008 $H_1$: $PaRA_M \neq PrRA_M$ Confirm $H_1$ | It shows that all of the designs as well as their average confirm that $PaRA_M \neq PrRA_M$ |
| Two-sample t-test for paired data (using mid-point CRRA's) | It tests if two variables have the same mean, assuming paired data (parametric) | $H_0$: The mean of the difference is zero | Reject $H_0$ at 10% Prob ($|T|>|t|$) = 0.0027 $PaRA_M \neq PrRA_M$ Confirm $H_1$ | Reject $H_0$ at 10% Prob ($|T|>|t|$) = 0.0005 $PaRA_M \neq PrRA_M$ Confirm $H_1$ | Reject $H_0$ at 10% Prob ($|T|>|t|$) = 0.0460 $PaRA_M \neq PrRA_M$ Confirm $H_1$ | Reject $H_0$ at 10% Prob ($|T|>|t|$) = 0.0005 $PaRA_M \neq PrRA_M$ Confirm $H_1$ | It shows that all of the designs as well as their average confirm that $PaRA_M \neq PrRA_M$ |

**Source(s):** Author's own work

**Table A5.** Comparative analysis of PaRA and PrRA across contexts: key findings from three statistical tests for *Women*

| Test | Test explanation | Test hypothesis | Women Task1 = Task6 (PaRA=PrRA) | Women Task2 = Task5 (PaRA=PrRA) | Women Task3 = Task4 (PaRA=PrRA) | Women PaRA=PrRA (average of Pa- and Pr-Tasks) | Overall conclusion |
|---|---|---|---|---|---|---|---|
| Wilcoxon matched-pairs signed-ranks test | It tests the equality of matched pairs of observations (non-parametric) | $H_0$: Both distributions are the same | Fail to Reject $H_0$ Prob>|z| = 0.9818 $H_1$: $PaRA_W \neq PrRA_W$ Cannot Confirm $H_1$ | Fail to Reject $H_0$ Prob>|z| = 0.3828 $H_1$: $PaRA_W \neq PrRA_W$ Cannot Confirm $H_1$ | Reject $H_0$ at 10% Prob>|z| = 0.0796 $H_1$: $PaRA_W \neq PrRA_W$ Confirm $H_1$ | Fail to Reject $H_0$ Prob>|z| = 0.4669 $H_1$: $PaRA_W \neq PrRA_W$ Cannot Confirm $H_1$ | Most designs as well as their average confirms that $PaRA_W$ and $PrRA_W$ are NOT statistically significantly different |
| Arbuthnott-Snedecor-Cochran sign test | It tests the equality of matched pairs of observations (non-parametric) | $H_0$: The median of the differences is zero (the true proportion of positive (negative) signs is one-half) | Fail to Reject $H_0$ Prob(.) = 1.0000 $H_1$: $PaRA_W \neq PrRA_W$ Cannot Confirm $H_1$ | Fail to Reject $H_0$ Prob(.) = 0.5716 $H_1$: $PaRA_W \neq PrRA_W$ Cannot Confirm $H_1$ | Fail to Reject $H_0$ Prob(.) = 0.1516 $H_1$: $PaRA_W \neq PrRA_W$ Cannot Confirm $H_1$ | Fail to Reject $H_0$ Prob(.) = 0.8601 $H_1$: $PaRA_W \neq PrRA_W$ Cannot Confirm $H_1$ | All designs as well as their average confirms that $PaRA_W$ and $PrRA_W$ are NOT statistically significantly different |
| Two-sample t-test for paired data (using mid-point CRRA's) | It tests if two variables have the same mean, assuming paired data (parametric) | $H_0$: The mean of the difference is zero | Fail to Reject $H_0$ Prob (|T|>|t|) = 0.7724 $H_1$: $PaRA_W \neq PrRA_W$ Cannot Confirm $H_1$ | Fail to Reject $H_0$ Prob (|T|>|t|) = 0.5466 $H_1$: $PaRA_W \neq PrRA_W$ Cannot Confirm $H_1$ | Reject $H_0$ at 10% Prob (|T|>|t|) = 0.0652 $H_1$: $PaRA_W \neq PrRA_W$ Confirm $H_1$ | Fail to Reject $H_0$ Prob (|T|>|t|) = 0.3736 $H_1$: $PaRA_W \neq PrRA_W$ Cannot Confirm $H_1$ | Most designs as well as their average confirm that $PaRA_W$ and $PrRA_W$ are NOT statistically significantly different |

**Source(s):** Author's own work

**Table A6.** Summary of statistical tests on payoff-risk premiums and price-risk premiums for *Men*

| Test | Test explanation | Test hypothesis | For men: RPTask2 = RPTask5 ($PaRP_M$ = $PrRP_M$) | Overall conclusion |
|---|---|---|---|---|
| Wilcoxon matched-pairs signed-ranks test | It tests the equality of matched pairs of observations (non-parametric) | $H_0$: Both distributions are the same | Reject $H_0$ Prob>|z| = 0.0013 $H_1$: $PaRP_M \neq PrRP_M$ Confirm $H_1$ | It shows that $PaRP_M \neq PrRP_M$ |
| Arbuthnott-Snedecor-Cochran sign test | It tests the equality of matched pairs of observations (non-parametric) | $H_0$: The median of the differences is zero (the true proportion of positive (negative) signs is one-half) | Reject $H_0$ Prob(.) = 0.0168 $H_1$: $PaRP_M < PrRP_M$ Confirm $H_1$ | It shows that $PaRP_M < PrRP_M$ |
| Two-sample *t*-test for paired data | It tests if two variables have the same mean, assuming paired data (parametric) | $H_0$: The mean of the difference is zero | Reject $H_0$ Prob(T < t) = 0.0002 $H_1$: $PaRP_M < PrRP_M$ Confirm $H_1$ | It shows that $PaRP_M < PrRP_M$ |

**Source(s):** Author's own work

**Table A7.** Summary of statistical tests on payoff-risk premiums and price-risk premiums for *Women*

| Test | Test explanation | Test hypothesis | For women: RPTask2 = RPTask5 ($PaRP_W$ = $PrRP_W$) | Overall conclusion |
|---|---|---|---|---|
| Wilcoxon matched-pairs signed-ranks test | It tests the equality of matched pairs of observations (non-parametric) | $H_0$: Both distributions are the same | Fail to reject $H_0$ Prob>|z| = 0.2588 $H_1$: $PaRP_W \neq PrRP_W$ Cannot Confirm $H_1$ | It shows that $PaRP_W$ and $PrRP_W$ are NOT statistically significantly different |
| Arbuthnott-Snedecor-Cochran sign test | It tests the equality of matched pairs of observations (non-parametric) | $H_0$: The median of the differences is zero (the true proportion of positive (negative) signs is one-half) | Fail to reject $H_0$ Prob(.) = 0.5716 $H_1$: $PaRP_W \neq PrRP_W$ Cannot Confirm $H_1$ | It shows that $PaRP_W$ and $PrRP_W$ are NOT statistically significantly different |
| Two-sample *t*-test for paired data | It tests if two variables have the same mean, assuming paired data (parametric) | $H_0$: The mean of the difference is zero | Fail to reject $H_0$ Prob(T < t) = 0.3627 $H_1$: $PaRP_W \neq PrRP_W$ Cannot Confirm $H_1$ | It shows that $PaRP_W$ and $PrRP_W$ are NOT statistically significantly different |

**Source(s):** Author's own work

**Note**

1. Payoff Risk Aversion (PaRA) refers to individuals' aversion to uncertainty in outcomes or final payoffs, typically measured through risky lottery choices based on the Direct Utility Function (DUF). In contrast, Price Risk Aversion (PrRA) captures individuals' aversion to uncertainty in the prices they face, elicited using the Indirect Utility Function (IUF). While both concepts fall under Expected Utility Theory, they differ in the domain of uncertainty—PaRA relates to what one receives, while PrRA concerns what one must pay.

2. The use of MCL designs is a well-established and widely accepted method for eliciting risk preferences in experimental economics (e.g. Holt and Laury, 2002; Binswanger, 1980). While each list captures only a single switching point, this point serves as a validated proxy for an individual's risk aversion under Expected Utility Theory and enables intuitive mapping to utility parameters such as CRRA. MCLs are particularly well-suited to this study because they allow for precise, tractable comparisons of risk preferences across price and payoff framings, while maintaining cognitive simplicity and incentive compatibility. By incorporating three distinct MCL designs under two framing conditions (DUF and IUF) and using a within-subject structure with six tasks per participant, the study enhances reliability and mitigates noise from any one task. Compared to alternative methods—which often impose higher cognitive burden and are vulnerable to anchoring and misreporting as discussed by Harrison *et al.* (2005) and Deck and Jahedi (2015)—MCLs provide a simple, incentive-compatible, and cognitively manageable format that aligns well with the goal of comparing the degrees of PaRA and PrRA across genders in a controlled, internally valid setting.

3. The use of the CRRA utility specification in this study is both theoretically grounded and methodologically appropriate for the context of the experimental tasks employed. CRRA is extensively used in the literature on experimental and applied economics due to its tractability, empirical plausibility, and alignment with expected utility theory when evaluating risky choices involving proportional or multiplicative changes in wealth (e.g. Holt and Laury, 2002; Gneezy and Potters, 1997). In contrast to Constant Absolute Risk Aversion (CARA), which assumes fixed risk aversion regardless of wealth levels, CRRA captures the empirically supported notion that individuals often exhibit decreasing absolute risk aversion and constant relative risk aversion in real-world economic decisions (Pratt, 1964; Arrow, 1971; Friend and Blume, 1975; Binswanger, 1980; Chiappori and Paiella, 2011). Moreover, given the lottery structures and price-versus-payoff uncertainties utilized in the experimental design, the CRRA framework provides a coherent and comparable measure of risk preferences across various tasks and different contexts utilized in this experiment.

4. Accordingly, they developed and calibrated their price-based MCL designs (i.e. tasks) by following the steps outlined by DT in economics (i.e. setting a budget constraint (with an endowment of $15), and deriving a Marshallian demand, and finding the IUF, and normalizing that IUF). They adopted and calibrated six equivalent risk elicitation designs in such a way that the six elicitation procedures were theoretically equivalent, given the EUT and DT. For the dataset collected and for more detailed information about the process of data collection, see Zeytoon-Nejad (2022) and Moosavian (2019).

5. Additionally, to minimize potential confounding influences, the order of tasks in each session and for each experimental subject was randomly assigned to address the "order effect" and "learning effect." The experimental design further accounts for several other potential sources of bias and variation, including the "incentive effect," "income effect," "wealth effect," "scale effect," "endowment effect," "selection effect," and "fixed effects," as elaborated in Zeytoon-Nejad (2022) and Moosavian (2019). These design considerations help ensure that any differences observed across the six elicitation tasks can be attributed to the phenomenon under investigation rather than to external or procedural artifacts, thereby enhancing the internal validity of the study.

6. The full experiment instructions set can be found at the following link: https://zeytoonnejad.wordpress.ncsu.edu/files/2022/07/Experiment-Instructions.pdf

7. Although this result is noteworthy, as it will be seen in what follows, this difference is not statistically significant in some of the statistical tests conducted and for some of the MCL designs used. The results of these statistical tests are reported in Appendix 2.

8. In the literature of experimental economics, a wide variety of other miscellaneous elicitation procedures have been used to elicit the degree of risk aversion in the lab. However, providing an exhaustive examination of these miscellaneous elicitation procedures is beyond the purview of this concise literature review. For a comprehensive list and detailed explanations of such miscellaneous procedures, interested readers are directed to resources such as Cox and Sadiraj (2008), Wilcox (2008), Harrison and Rutström (2009), and Charness *et al.* (2013).

**Further reading**

**Corresponding author**

Ali Zeytoon-Nejad can be contacted at: zeytoosa@wfu.edu